\documentclass[final]{aipproc}
\layoutstyle{6x9}

\usepackage{amsfonts}
\usepackage{amsmath}
\usepackage{amssymb}

\newcommand{\ar}{\arrowvert}

\newcommand{\cd}{\! \cdot \!}
\newcommand{\be}{\begin{equation}}
\newcommand{\ee}{\end{equation}}
\newcommand{\ba}{\begin{eqnarray}}
\newcommand{\ea}{\end{eqnarray}}

\newcommand{\bs}{\bf}
\begin{document}

\title{Franck-Condon principle for  heavy-quark  hadron decays}


\author{Felipe  Llanes-Estrada}{
  address={Depto. F\'isica Te\'orica I, Universidad Complutense de Madrid,
28040 Madrid, Spain}
}

\author{Steve  Cotanch}{
  address={Department of Physics, North Carolina State University, 
 Raleigh NC  27695, USA}
}

\author{Ignacio General}{
  address={Bayer School of Natural  and
Environmental  Sciences,
Duquesne University, Pittsburgh, PA 15282, USA}
}

\author{Ping Wang}{
  address={Jefferson Laboratory, 12000 Jefferson Ave., Newport News, VA 23606, USA
}
}

\begin{abstract}
The Franck-Condon principle governing molecular 
electronic transitions is utilized to study heavy-quark hadron decays.
This provides
a direct assessment of the wavefunction of the parent hadron  
if the momentum distribution of the open-flavor decay products is measured.
Model-independent results include an experimental distinction between  quarkonium
and  exotica (hybrids, tetraquarks...), an off-plane 
correlator  signature for tetraquarks and a direct
probe of the  sea quark orbital wavefunction
relevant in the discussion of
 $ ^3S_1$ or $ ^3P_0$ decay mechanisms.
\end{abstract}

\maketitle

\section{The Franck-Condon principle}
Fluorescence spectra were first explained in 1925 by James Franck and Edward 
Condon on the basis of what is now known as the Franck-Condon [FC] principle.
Recognizing the slow nuclear degrees of freedom, they asserted that for any molecular electronic transition (absorption or emission)  there is no appreciable change in the internuclear coordinate separation or momentum 
(the scale separation between the electron mass
and the nuclear mass being at least $10^{-3}$). 
We propose that this principle can also be used
to experimentally obtain the momentum distribution in 
heavy-quark systems. Our conjecture is that \emph{the heavy-quark momentum
distribution in the decaying hadron coincides with
the momentum distribution of the decay-product hadrons each carrying a heavy quark.}
That is, up to corrections of order $\Lambda_{QCD}/M_Q$ (alternatively,
$\alpha_s$ for ground state heavy quarkonium), and if
$E/p$ conservation permits, a measurement of the decay hadrons
yields  information about the parent hadron's  wavefunction, thus allowing useful tests for exoticness, structure of the 
Fermi sea and other applications. 

\section{Quarkonium versus hybrid signature} 

With the $\psi(4260)$, $\psi(4320)$ and $\psi(4620)$ recently discovered at  
$B$-factories and the  $\psi(4040)$, $\psi(4160)$ and 
$\psi(4400)$ observed in the total cross section 
$e^-e^+\to {\rm hadrons}$,
there is now a clear excess of expected states  in the charmonium spectrum. 
This overpopulation is explained in many-body 
approaches \cite{oldhybrids,General:2006ed,General:2007bk,Buisseret:2007ed} which  
predict several $c\bar{c}g$ hybrid mesons 
in addition to the $c\bar{c}$  spectrum.  
where four additional states are predicted in that 
The challenge for  theorists is to identify and distinguish such  
states. To this end we propose using the model-independent 
FC principle to extract the parton probability distribution from 
experiment. 

\begin{figure}[h]
\includegraphics[height=.28\textheight,angle=-90]{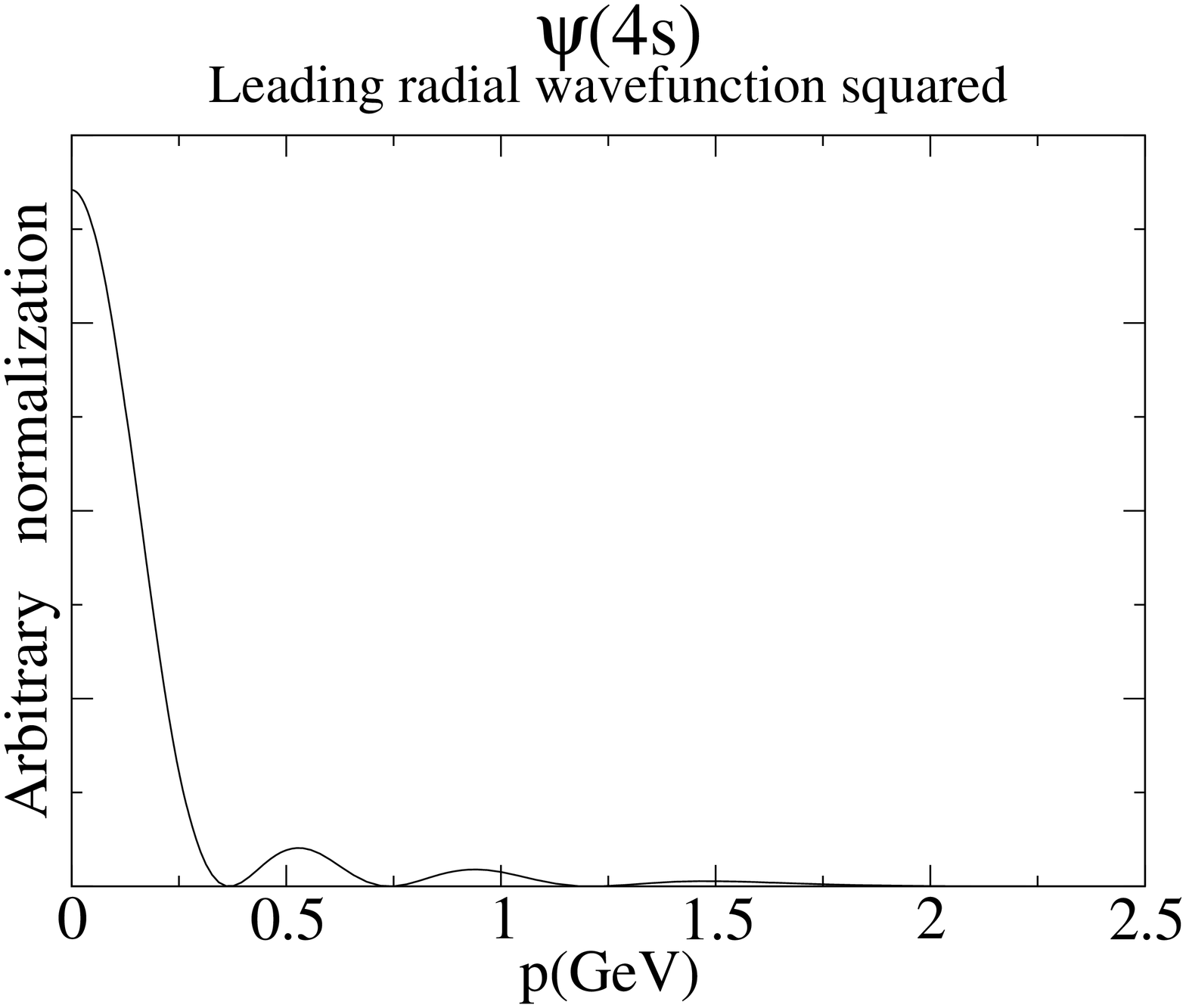}
\includegraphics[height=.28\textheight,angle=-90]{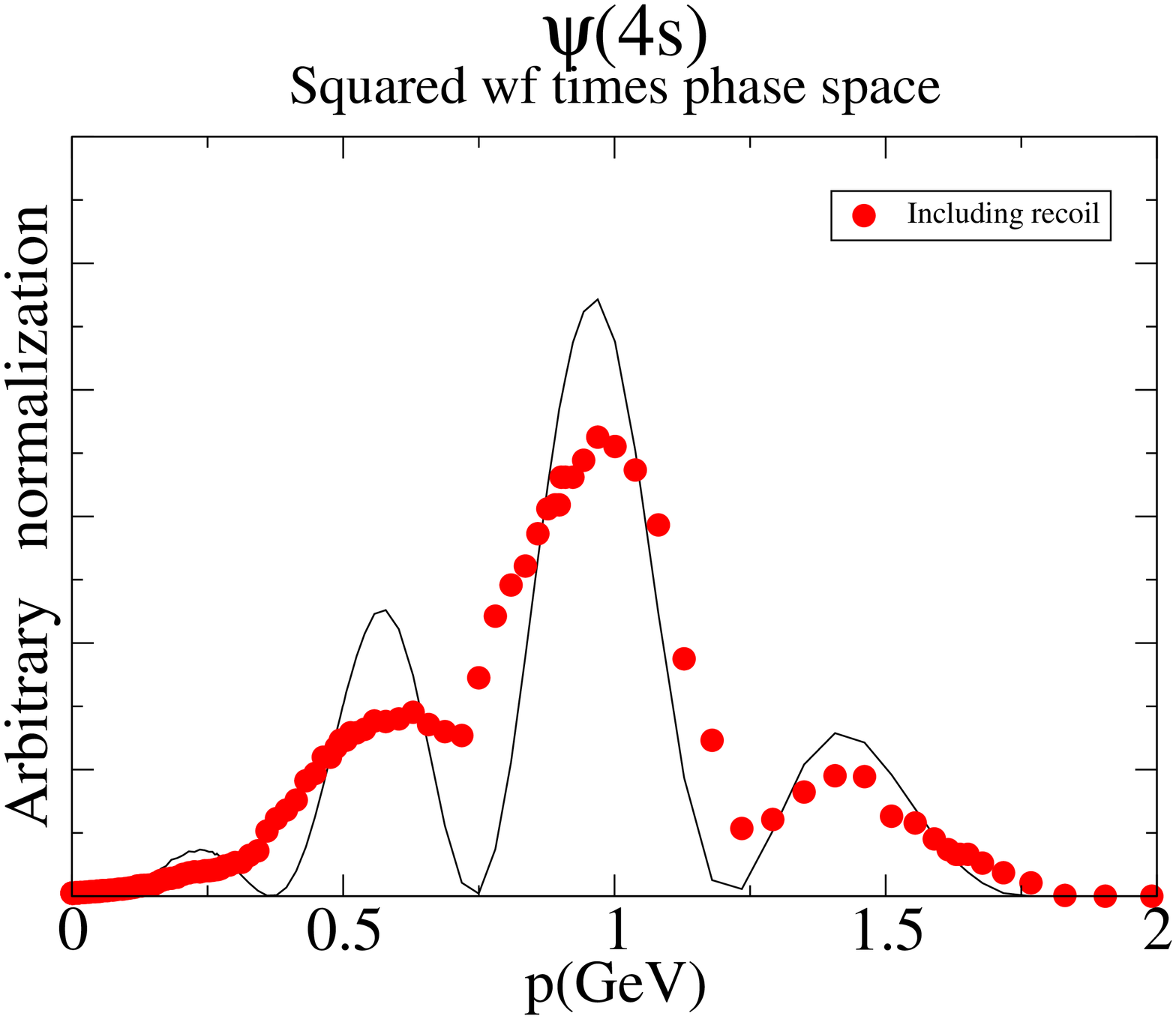}
\caption{Left: the 4S  quarkonium wavefunction in the 
Coulomb gauge model \cite{oldhybrids}. 
Right: the  momentum distribution for the $D\bar{D}$ mesons in a 
three-body $D\bar{D}\pi$ decay, using the Franck-Condon 
principle for $c\bar{c}\to D\bar{D}$ (solid line) and momentum smearing 
(red dots).  Note the preservation of wavefunction nodes.
\label{fig:4s}}
\end{figure}

Consider the decay of the excited charmonium state, $\psi(4S)$ to $D\bar{D} \pi$.
We focus on a three-body decay since in a two-body decay momentum
conservation precludes using the FC principle.
The momentum distribution of the $c$ quarks is plotted in the first graph of Fig. \ref{fig:4s}.
Invoking the  FC conjecture, the momentum distribution of the decay $D$ mesons
is given by the second plot, now versus $\ar {\bf p}_D- {\bf p}_{\bar{D}} \ar$, in Fig. \ref{fig:4s} which incorporates phase space and recoil effects from a finite quark mass.  
Note that
after accounting for momentum 
smearing of order 150-200 MeV,
 the
3 Sturm-Liouville nodes, which must be preserved, are clearly
depicted as dips in this spectrum.

\begin{figure}[h]
\includegraphics[height=.28\textheight,angle=-90]{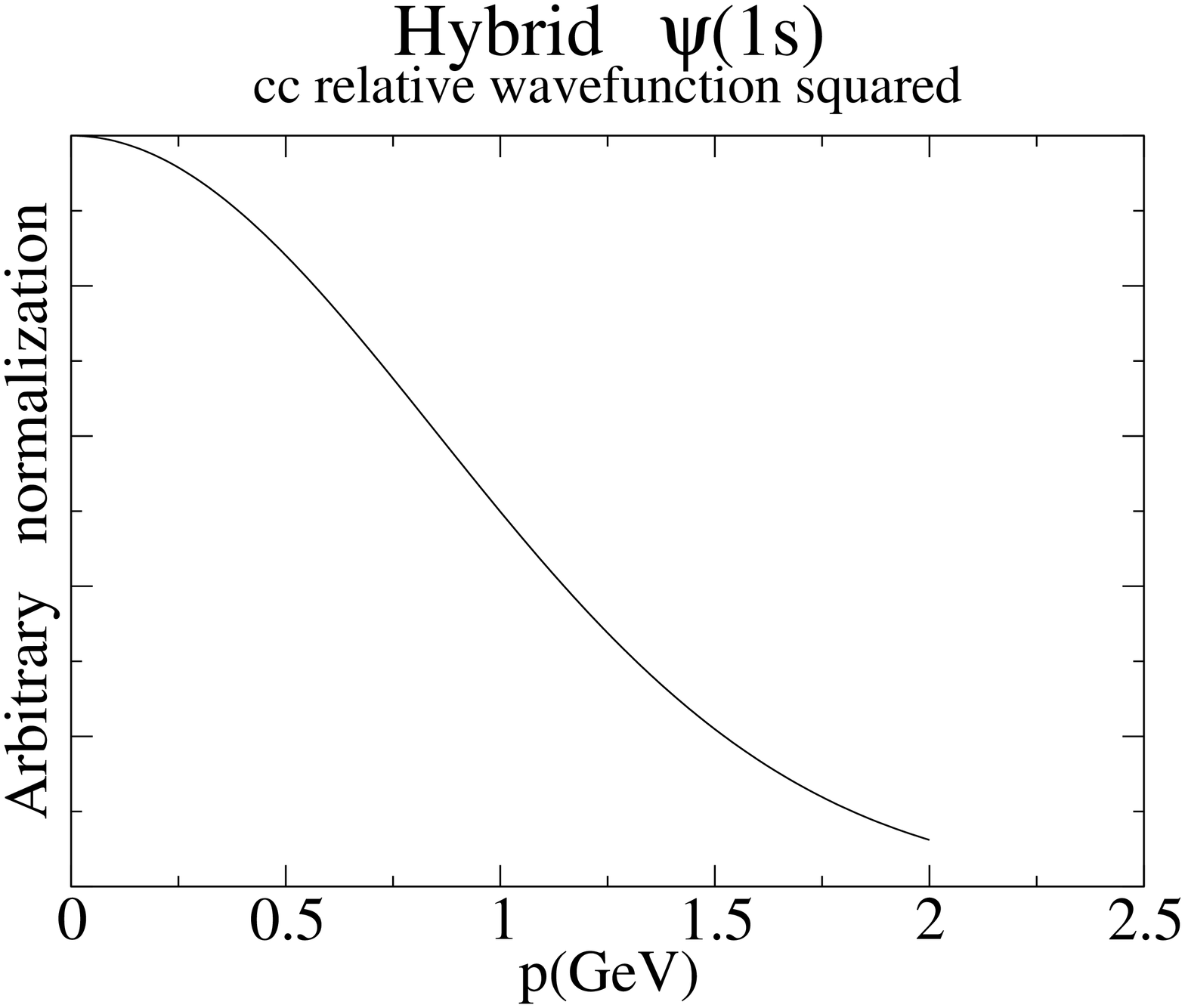}
\includegraphics[height=.28\textheight,angle=-90]{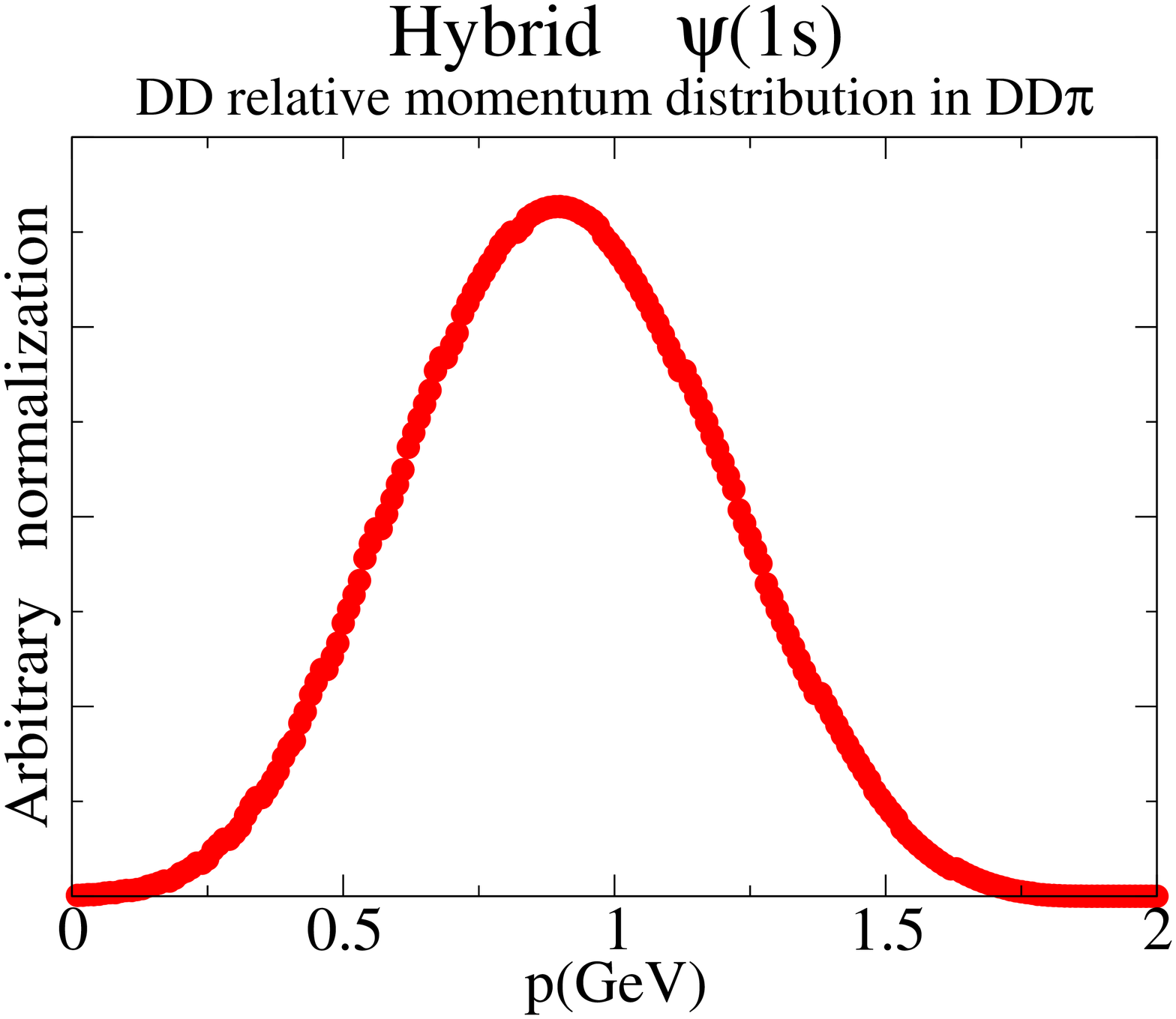}
\caption{As in Fig. \ref{fig:4s}  for the ground state $c \bar{c}g$  hybrid 
meson with comparable mass in the 4.2-4.5 GeV region.
Note  the 
wavefunction is  nodeless and the final relative momentum 
distribution of the $D\bar{D}$ pair has a simple bell-shape with no shoulders.
 \label{fig:1s}}
\end{figure}

These distinctive dips are to be contrasted with the predicted \cite{oldhybrids,General:2006ed} bell-shaped momentum distribution 
 in Fig. \ref{fig:1s} that corresponds to a ground state $c\bar{c}g$ hybrid 
meson with comparable mass. The  gluonic excitation is 
responsible for the additional energy that places this state high in the 
charmonium spectrum. The  relative $c \bar{c}$ wavefunction in the hybrid state  is  nodeless and the 
corresponding decay spectrum is much simpler than in the case of an excited 
$c\bar{c}$.
Hence by 
measuring the relative momentum of the $D$ mesons in the $D\bar{D}\pi$ 
subsystem, 
 a model-independent way of distinguishing 
 between a hybrid and quarkonium appears possible.
 Again note that $D\bar{D}$ decay alone is not useful since the two mesons  
momentum distribution is a Dirac delta function, which invalidates applying the
FC principle.
Related, $D\bar{D}^*$ decays also contribute to this three-body 
channel from a two-body decay which again avoids the FC 
constraint and need to be excluded with a kinematic cut in the Dalitz 
plot for $D\bar{D}\pi$.

Although the Belle collaboration has  reported 
$D\bar{D}K$ spectra \cite{:2007aa} stemming from weak B decays, this 
cannot be  used for our  studying since recoil 
corrections are larger for a kaon. We therefore await 
$D\bar{D}\pi$ data.
Fortunately, the $b \bar{b}$  system provides an even more favorable application of the FC principle
and
 the Belle collaboration has collected $20\ fb^{-1}$ of 
integrated luminosity for the $\Upsilon(5S)$. This state, at $10860\ MeV$, has sufficient phase 
space to decay to $B\bar{B}\pi$ and can therefore be used to repeat the 
above analysis. This state's position is quite well 
reproduced by $b\bar{b}$ bottomonium models, and therefore its nature as 
(largely) a bottomonium excitation has not been questioned.
The FC principle can be considered validated if the spectra of 
$B\bar{B}$ relative momenta   
in the $B\bar{B}\pi$  center of mass frame has visible shoulders. 

\section{Off-plane correlator: tetraquark signature}

From several studies \cite{General:2006ed,General:2007bk,Buisseret:2007ed} one can conclude that 
ground state tetraquarks and hybrids in the charmonium region can coexist 
with similar mass, and therefore the Sturm-Liouville nodal difference will not be useful.
However, one can exploit the fact that a three-body hybrid meson has a planar structure
with two 
independent momenta. The same structure emerges in the flux tube model, where the flux tube mode carries one unit
of transverse excitation.
In contrast, tetraquark systems have three independent momenta
with one  out of plane even in the center of momentum frame.
Applying the Franck-Condon principle we deduce that a measurement of the 
off-planarity of the four-meson decay channel of the parent hadron 
measures the off-planarity of the latter in terms of its intrinsic 
wavefunction. 
Therefore one has to look for four-meson decay channels such as 
$D\bar{D}K\bar{K}$, and study their off-planarity in the center of momentum
frame.

A measure of the deviation from planarity of a four-meson system has been presented
in Ref. \cite{General:2007bk} where the 
off-plane correlator
$$
\Pi({\bs p}_1,{\bs p}_2,{\bs p}_3,{\bs p}_4) =
\frac{( ({\bs p}_1\times {\bs p}_2) \cd {\bs p}_3)^2}{\sqrt{ \ar
{\bs p}_1 \times {\bs p}_2 \ar \ar {\bs p}_2 \times {\bs p}_3 \ar
\ar {\bs p}_1 \times {\bs p}_3 \ar \ar {\bs p}_1 \times {\bs p}_4
\ar \ar {\bs p}_2 \times {\bs p}_4 \ar \ar {\bs p}_3 \times {\bs
p}_4 \ar
}} 
$$
was introduced.
The 
${\bf p}_{i}$ form a parallelepiped whose volume is zero if any three lie 
on the same plane. Therefore the volume $|{\bf p}_3\cdot({\bf p}_1\times {\bf p}_2)|$ measures
the deviation from coplanarity. However, this volume is also proportional to the
length of the sides, i.e. depends on the phase space for the decay of the
parent hadron and not only on its internal structure. The normalized correlator is a dimensionless, 
pure number, and therefore independent of the phase-space.
It is also invariant under  permutations  and 
is zero if the Franck-Condon principle exactly holds for the 
decay of a meson with a two or three-body internal wavefunction.
The maximum value of $\Pi\simeq 0.707$ seems to be attainable by a
symmetric tetrahedral configuration of all 4 momenta, as we have checked
by Montecarlo simulations. A typical 
high value for this correlator is 0.59, corresponding to three of the
mesons having equal momentum at right angles (along the edges of a cube), with
the fourth momentum ${\bf p}_4=-{\bf p}_1-{\bf p}_2-{\bf p}_3$  longer by a factor 
$\sqrt{3}$. 

\begin{figure}[h]
\includegraphics[height=.3\textheight,angle=-90]{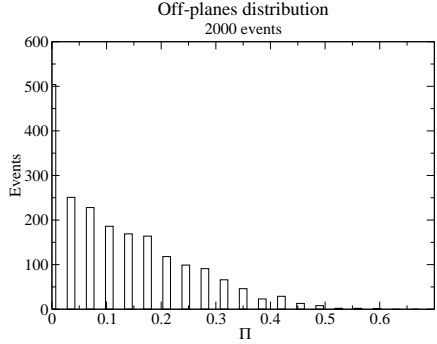}
\caption{Distribution for a sample of 2000 four-meson events. The 
Montecarlo sample has been generated using a tetraquark model wave-function subject to 
the Franck-Condon constraint. 
\label{fig:offplanes}
}
\end{figure}

As an  example from current meson studies, we have analyzed the 
decay of the newly found $Z(4430)$ meson, observed by the Belle 
collaboration in the $\psi'\pi$  spectrum for the decay $B \to KZ \to K\psi'\pi$.
The flavor quantum numbers of the $Z$ meson are those of a $\pi$ or $\rho$,
namely $u\bar{d}$ for the $Z^+$. However, its narrow width (40-50 MeV) for such
a heavy meson, as well as its discovery decay mode, suggest that it has
 a significant component that is hidden-exotic, $u\bar{d}c\bar{c}$.
Therefore having sufficient phase-space, it may decay  
$Z  \to D\bar{D} \pi\pi$. Although in this example not all mesons have
heavy quarks necessary for a rigorous application of the FC principle,
it is still worthwhile to examine the distribution
presented in Fig. \ref{fig:offplanes}.

The calculation in the figure is the distribution of the off-plane 
correlator for a random sample of 2000 events, taking the four 
three-momenta of the quarks (that coincide exactly with the four 
meson-momenta only if the Franck-Condon principle exactly holds). The 
quark momenta  have been generated with a standard pseudorandom 
number generator and distributed according to a  model Gaussian 
wavefunction for the tetraquark,
$\psi(p_1,p_2,p_3)= {\mathcal N}(p_I/\alpha_I) exp(-\frac{p^2_A}{\alpha^2_A} -
\frac{p^2_B}{\alpha^2_B} - \frac{p^2_I}{\alpha^2_I})$,
with ${\bf p}_A={\bf p}_1-{\bf p}_2$, ${\bf p}_B={\bf p}_3-{\bf p}_4$ the 
cluster relative momenta and ${\bf p}_I={\bf p}_1+{\bf p}_2-{\bf p}_3-{\bf p}_4$ the intercluster momenta.  The parameters are
 $\alpha_A=\alpha_B=1.35 \ GeV$, $\alpha_I=0.6\ GeV$.  See  Ref.  \cite{General:2007bk}
 for a complete discussion of the variational tetraquark wavefunction.
 As can be seen, the off-plane correlator is peaked at low values but 
is well-populated in the region 0.2-0.4 and even approaches 
the value found for the cubic configuration, 0.59.  Again, even though the FC principle does
not rigorously hold in this example, we submit that the non-zero correlator value is a
representative result for  tetraquark structure, quite distinct from the expected value
of zero for a hybrid wavefunction.

It is interesting  that a paper  at this 
workshop by  P. Bicudo and M. Cardoso also
proposes that the $Z(4430)$ is a $D^*D_1$ molecule-type tetraquark decaying to $\psi'\pi^+$ and
that the two $D$ mesons have
a node in their relative wavefunction. They obtain widths for this meson
of order $0.2\ MeV$ for the channel $J/\psi \pi^+$ and $4.6\ MeV$ for 
the observed $\psi'\pi^+$, through the application
of Moshinsky-Ribeiro-Van Beveren wavefunction overlap rules.  However, this is also 
equivalent to the Franck-Condon principle in reverse, where the 
original tetraquark meson has a wavefunction  similar
to the $\psi'$ rather than the $J/\psi$ as revealed by the decay pattern.

\section{Spin of sea pairs: distinguishing decay mechanisms}

\begin{figure}[h]
\includegraphics[height=.2\textheight]{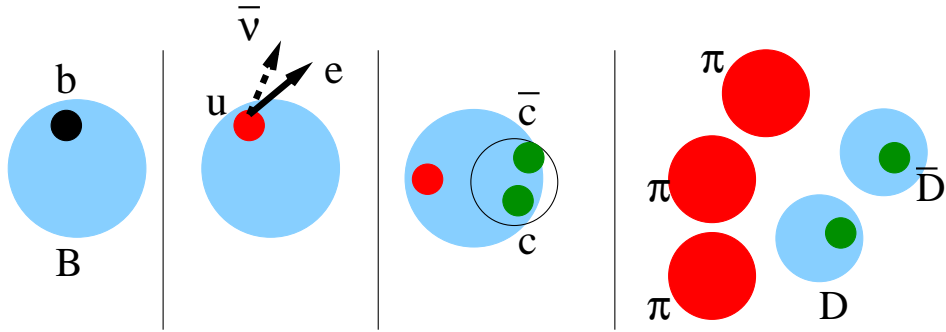}
\caption{From left to right:
a $b$ quark in a $B$ meson undergoes a CKM-suppressed weak
decay. The excited hadron  state decays by $c\bar{c}$ formation with
relative orbital wavefunction to be determined.
The final hadron decay products are two $D$ mesons
and an arbitrary number of pions. Whether the two $D$ mesons are in a 
relative $s$ or $p$-wave determines the corresponding angular 
wavefunction of the original $c\bar{c}$ pair by the 
Franck-Condon principle. \label{fig:Bdecay}}
\end{figure}

An unanswered question in hadron physics is whether the spins of quark-antiquark
pairs from the hadronic Fermi sea  are  correlated. While  deep inelastic scattering data \cite{wagner} reveals that the 
sea quarks  are largely unpolarized, it has not been used to
extract quark-antiquark correlations.
This question is important to understand meson decays since the valence quark
component of the parent and daughter  hadrons can be different which would require
a pair creation transition. 
In QCD, quarks couple to negative parity vector gluons
suggesting  the $q\bar{q}$ pair
are in a $ ^3S_1$ $s$-wave. However, extensive quark model phenomenology
\cite{barnes}  indicates  a $ ^3P_0$ scalar wavefunction.
We argue here that the Franck-Condon principle can resolve this 
point in a model-independent way by extracting the correlation of a
$c\bar{c}$ vacuum pair in a heavy-quark hadron decay. Such analysis would require
a pair of $D\bar{D}$ mesons, and this favors $B$ 
decays, since the $B$ meson is heavy enough to decay to two $D$ mesons,
 and copious samples, of order $500$ million, have been produced at
the $B$-factories. 
The channel we recommend for the analysis is the semileptonic decay
$
B \to e\bar{\nu}\  D\bar{D} \ n\pi
$ 
where the two $D$ mesons are accompanied by an electron and $n$
pions as depicted in Fig. \ref{fig:Bdecay}.

The semileptonic decays only account for about $10\%$ of the total $B$ samples
and it is necessary to trigger with a fast electron to avoid  charm quark counts
from weak $b$ decays, $b\to c\bar{c}s$ or $b\to c\bar{c}d$, that
are a background to the strong charmed decays. Decays with one kaon or
only one $D$ meson are likewise to be discarded as a charmed quark could
come from this weak decay. 
Unlike the previous application, 
$D^*\bar{D}$ and $D^*\bar{D}^*$ can now be used in place of $D\bar{D}$.
The observable is the relative orbital wavefunction of the 
$D\bar{D}$ mesons, accompanied by any number of pions that balance energy-momentum. The $D$ meson reconstruction efficiency is very low at  
$B$-factories, around $1\%$, and the $B$ meson decays to $D$ pairs also have  a small
branching fraction of $1\%$ respect to similar light-light decays \cite{pdg}.
We thus expect samples of order $10^{-5}$ of the total number of
events. 
It is not necessary to reconstruct the $B$ meson decay, so
the $b$ quark may be tagged by the other $B$ meson in the 
$Y(4S)\to B\bar{B}$ reaction. The two $D$ mesons 
however need to be fully reconstructed and their momentum measured 
accurately enough to perform an angular analysis to distinguish 
between a $s$ or $p$ wave  relative wavefunction.

\section{Summary}
We have shown that the Franck-Condon
principle can be an effective tool to experimentally extract
information about the wavefunction in
heavy quark systems, especially for documenting exotic degrees of freedom.
We believe  further applications of this constraint in high statistics measurements
will provide significant hadron structure insight.


\begin{theacknowledgments}
  We thank the organizers of Scadron70 for  a 
very successful meeting. Work supported by grants DOE DE-FG02-03ER41260, BSCH-PR34/07-15875, 
FPA 2004 02602, FPA 2005-02327
and Acci\'on Integrada Hispano-Portuguesa HP2006-0018.
\end{theacknowledgments}

\end{document}